\documentstyle[amssymb,12pt]{article}

\begin{document}

\title{ALL POISSON STRUCTURES IN~${\mathbb R}^3$}
\author{ Metin G{\" u}rses and Kostyantyn Zheltukhin\\
{\small Department of Mathematics, Faculty of Sciences}\\
{\small Bilkent University, 06800 Ankara, Turkey}\\}

\begin{titlepage}
\maketitle

\begin{abstract}
Hamiltonian formulation of $N=3$ systems is considered in general.
The most general solution of the Jacobi equation in ${\mathbb
R}^3$ is proposed. Compatible Poisson structures and the
corresponding bi-Hamiltonian $N=3$ systems are also discussed.
\end{abstract}

\end{titlepage}

\section*{1. Introduction.}

Hamiltonian formulation of $N=3$ systems has been intensively
considered in the last two decades. Recent works \cite{ber1},
\cite{ber2} on this subject give a very large class of the
solutions of the Jacobi equation for the Poisson matrix $J$. Very
recently generalizing the solutions given in \cite{ber1} we
proposed new classes of solutions of the Jacobi equation and gave
also some compatible Poisson structures. In the same work we have
considered also bi-Hamiltonian systems corresponding to compatible
Poisson structures \cite{gur}.

In this work we give the most general solution of the Jacobi
equation in ${\mathbb R}^3$, and give the condition for the
existence of the compatible pairs of Poisson structures. Finally
we give the conditions for the existence of the bi-Hamiltonian
systems.

Matrix $J=(J^{ij}), \quad i,j=1,2,3,$ defines a Poisson structure
in ${\mathbb R}^3$ if it is skew-symmetric, $J_{ij}=-J_{ji}$, and
its entries satisfy the Jacobi equation
\begin{equation}\label{jacobi1}
J^{li}\partial_l\,J^{jk}+J^{lj}\partial_l\,J^{ki}+J^{lk}\partial_l\,J^{ij}=0,
\end{equation}
where $i,j,k=1,2,3$. Here we use the summation convention, meaning
that repeated indices are summed up. We show that the general
solution of the above equation (\ref{jacobi1}) has the following
form
\begin{equation}\label{GenSol}
J^{ij}=\mu\epsilon^{ijk}\partial_k\Psi,
\end{equation}
where $\mu$ and $\Psi$ are arbitrary differentiable functions of
$x^{i}$, $i=1,2,3$ and $\epsilon^{ijk}$ is the Levi-Civita symbol.
This has a very natural geometrical explanation. Let $\Psi=c_{1}$
and $H=c_{2}$ define two surfaces $S_{1}$ and $S_{2}$
respectively,in ${\mathbb R}^3$ , where $c_{1}$ and $c_{2}$ are
some constants. Then the intersection of these surfaces define a
curve $C$ in ${\mathbb R}^3$. The velocity vector ${dx \over dt}$
of this curve is parallel to the vector product of the normal
vectors $\nabla \Psi$ and $\nabla H$ of the surfaces $S_{1}$ and
$S_{2}$, respectively, i.e.,

\begin{equation}
{dx \over dt}=-\mu\, \nabla \Psi \times \nabla H, \label{sur}
\end{equation}

\noindent where $\mu$ is any arbitrary function in ${\mathbb
R}^2$. This equation defines a Hamiltonian system in ${\mathbb
R}^3$. We shall prove that all Hamiltonian systems in ${\mathbb
R}^3$ are of the form given in (\ref{sur}).

As we shall see in Section 3 and 4, having the Poisson structure
in the form (\ref{GenSol}) allows us to construct compatible
Poisson structures $J$ and $\tilde{J}$ and hence the corresponding
bi-Hamiltonian systems. For further details please see \cite{gur}.

\section*{2. General Solution.}
Let $J^{12}=u$, $J^{13}=-v$, $J^{23}=w$ , where $u, v $ and $w$
are some differentiable functions of $x_{1}, x_{2}$ and $x_{3}$.
Then Jacobi equation (\ref{jacobi1}) takes the form
\begin{equation}\label{jacobi2}
u\partial_1v - v\partial_1u + w\partial_2u -w\partial_2u +
v\partial_3w-w\partial_3v=0.
\end{equation}
Assuming that $u \ne 0$, let $\displaystyle{ \rho=\frac{v}{u}}$
and $\displaystyle{ \chi=\frac{w}{u}}$ then equation
(\ref{jacobi2}) can be written as
\begin{equation}\label{jactr1}
\partial_1\rho-\partial_2\chi+\rho\partial_3\chi-\chi\partial_3\rho=0.
\end{equation}
This equation can be put in a more suitable form by writing it
\begin{equation}\label{jactr2}
(\partial_1-\chi\partial_3)\rho-(\partial_2-\rho\partial_3)\chi=0.
\end{equation}
Introducing differential operators $D_1$ and $D_2$ defined by
\begin{equation}
D_1=\partial_{1}-\chi\partial_{3} \qquad
D_2=\partial_{2}-\rho\partial_{3},
\end{equation}
one can write equation (\ref{jactr2}) as
\begin{equation}
D_1\rho-D_2\chi=0. \label{eq1}
\end{equation}

\vspace{0.3cm}

\noindent {\bf Lemma 1.} {\it Let equation (\ref{eq1}) be
satisfied. Then there are new coordinates $\bar{x}_1, \bar{x}_2,
\bar{x}_3$ such that
\begin{equation}\label{nD}
D_1=\partial_{\bar{x}_1},\quad {\mbox and} \quad
D_2=\partial_{\bar{x}_2}.
\end{equation}}

\vspace{0.3cm}

\noindent {\bf Proof.} If equation (\ref{eq1}) is satisfied, it is
easy to show that the operators $D_{1}$ and $D_{2}$ commute, i.e.,
$$
D_1\circ D_2-D_2\circ D_1=0.
$$
Hence, by the Frobenius theorem (see \cite{Olv} p. 40) there exist
coordinates $\bar{x}_1, \bar{x}_2, \bar{x}_3$ such that the
equalities (\ref{nD}) hold. $\Box$

\vspace{0.3cm}

\noindent The coordinates $\bar{x}_1, \bar{x}_2, \bar{x}_3$ are
described by the following lemma.

\vspace{0.3cm}

\noindent {\bf Lemma 2.} {\it Let $\zeta$ be a common invariant
function of $D_1$ and $D_2$, i.e.
\begin{equation}
D_1\zeta=D_2\zeta=0 \label{eq2},
\end{equation}
then the coordinates $\bar{x}_1, \bar{x}_2, \bar{x}_3$ of Lemma 1
are given by
\begin{equation}
\bar{x}_1=x_1,\quad \bar{x}_2=x_2,\quad \bar{x}_3=\zeta.
\end{equation}
Moreover from (\ref{eq2}) we get ($\partial_{3} \zeta \ne 0$),
\begin{equation}\label{rhochi}
\rho=\frac{\partial_1\zeta}{\partial_3\zeta},\quad
\chi=\frac{\partial_2\zeta}{\partial_3\zeta}.
\end{equation}
}

\vspace{0.3cm}

\noindent {\bf Theorem 1.} {\it All Poisson structures in
${\mathbb R}^3$ take the form (\ref{GenSol}), i.e., $J^{ij}=\mu\,
\epsilon^{ijk}\,
\partial_{k}\, \zeta$. Here $\mu$ and $\zeta$ are some
differentiable functions in ${\mathbb R}^3$ }

\vspace{0.3cm}

\noindent {\bf Proof.} Using (\ref{rhochi}), the entries of matrix
$J$, in the coordinates $\bar{x}_1, \bar{x}_2, \bar{x}_3$, can be
written as
\begin{equation}
\begin{array}{llr}
u&=& \mu\partial_3\zeta,\\
v&=& \mu\partial_2\zeta,\\
w&=& \mu\partial_1\zeta.\\
\end{array}
\end{equation}
Thus matrix $J$ has the form (\ref{GenSol})\, ($\Psi=\zeta$).
$\Box$

\vspace{0.3cm}

\noindent {\bf Remark 1}.\, So far we assumed that $u \ne 0$. If
$u=0$ then the Jacobi equation becomes quite simpler $v
\partial_{3}\, w- w \partial_{3}\, v=0$ which has the simple
solution $w= v \, \xi(x_{1}, x_{2})$, where $\xi$ is an arbitrary
differentiable of $x_{1}$ and $x_{2}$. This class is also covered
in our general solution (\ref{GenSol}) by letting $\Psi$
independent of $x_{3}$.

\vspace{0.3cm}

\noindent A well known example of a dynamical system with Poisson
structure of the form (\ref{GenSol}) is the Euler equations.

\vspace{0.3cm}

\noindent {\bf Example 1.} Consider the Euler equations
(\cite{Olv}, pp.397--398,)
\begin{equation}\label{e}
\begin{array}{lll}
\dot{x}_1&=&\displaystyle{\frac{I_2-I_3}{I_2I_3}}x_2x_3,\\
\dot{x}_2&=&\displaystyle{\frac{I_3-I_1}{I_3I_1}}x_3x_1,\\
\dot{x}_3&=&\displaystyle{\frac{I_1-I_2}{I_1I_2}}x_1x_2,\\
\end{array}
\end{equation}
where $ I_1,I_2,I_3\in {\mathbb R}$ are some (non-vanishing) real
constants. This system admits Hamiltonian representation of the
form~(\ref{GenSol}). The matrix $J$ can be defined in terms of
function $\Psi=-\frac{1}{2}(x^2_1+x^2_2+x^2_3)$ and $\mu=1$, so
\begin{equation}
\begin{array}{lll}
u&=&-x_3,\\
v&=&-x_2,\\
w&=&-x_1,\\
\end{array}
\end{equation}
and
$H=\displaystyle{\frac{x^2_1}{2I_1}+\frac{x^2_2}{2I_2}+\frac{x^2_3}{2I_3}}$.\\

\vspace{0.3cm}

\noindent {\bf Remark 2.} All of the Poisson structures described
in \cite{ber1} and \cite{gur} have the form (\ref{GenSol}). The
Poisson structure $J$, described in these references are given by
\begin{equation}\label{cl1}
\begin{array}{rrr}
u(x)&=&\eta(x_1,x_2,x_3)\psi_1(x_1)\psi_2(x_2)\phi_3(x_3),\\
v(x)&=&\eta(x_1,x_2,x_3)\psi_1(x_1)\phi_2(x_2)\psi_3(x_3),\\
w(x)&=&\eta(x_1,x_2,x_3)\phi_1(x_1)\psi_2(x_2)\psi_3(x_3),\\
\end{array}
\end{equation}
where ${\eta(x_1,x_2,x_3),\quad \psi_i(x_i),\quad \phi_i(x_i)},
\quad i=1,2,3,$ are arbitrary non vanishing differentiable
functions. Defining
$\mu=\eta(x_1,x_2,x_3)\psi_1(x_1)\psi_2(x_2)\psi_3(x_3)$ and
$\Psi=\displaystyle{ \int^{x_{1}}
\frac{\phi_1}{\psi_1}dx_1-\int^{x_{2}}
\frac{\phi_2}{\psi_2}dx_2}+\int^{x_{3}} \frac{\phi_3}{\psi_3}dx_3$
one obtains that $J$ has form (\ref{GenSol}).

\vspace{0.5cm}

\noindent {\bf Remark 3}.\,In general the Darboux theorem states
that (see \cite{Olv}) , locally, all Poisson structures can be
reduced to the standard one (Poisson structure with constant
entries). The above theorem , Theorem 1 , resembles the Darboux
theorem for $N=3$. All Poisson structures, at least locally, can
be cast into the form (\ref{GenSol}). This result is important
because the Dorboux theorem is not suitable for obtaining
multi-Hamiltonian systems in ${\mathbb R}^3$, but we will show
that our theorem is effective for this purpose. Writing the
Poisson structure in the form (\ref{GenSol}) allows us to
construct bi-Hamiltonian representations of a given Hamiltonian
system.

\vspace{0.3cm}

\noindent {\bf Definition 1.} {\it Two Hamiltonian matrices $J$
and $\tilde{J}$ are compatible, if the sum $J+\tilde{J}$ defines
also a Poisson structure.}

\vspace{0.3cm}

\noindent {\bf Lemma 3.} {\it Let Poisson structures $J$ and
$\tilde{J}$ have form (\ref{GenSol}), so
$J^{ij}=\mu\epsilon^{ijk}\partial_k\Psi$ and
$\tilde{J}^{ij}=\tilde{\mu}\epsilon^{ijk}\partial_k\tilde{\Psi}$.
Then $J$ and $\tilde{J}$ are compatible if and only if there exist
a differentiable function $\Phi(\Psi,\tilde{\Psi})$ such that
\begin{equation}
\tilde{\mu}=\mu\frac{\partial_{\tilde{\Psi}}\Phi}{\partial_{\Psi}\Phi},
\label{con}
\end{equation}
provided that $\partial_{\Psi}\, \Phi \equiv {\partial \Phi \over
\partial \Psi} \ne 0$ and $\partial_{\tilde{\Psi}}\, \Phi \equiv
{\partial \Phi \over \partial \tilde{\Psi}} \ne 0$

}

\vspace{0.3cm}

\noindent {\bf Remark 4}.\, In \cite{ber2} a method is given to
find new solutions of the Jacobi equation (\ref{jacobi1}) from the
known ones. Construction of compatible structures, Lemma 3, covers
all such cases. Letting, for instance,
$\tilde{\Psi}=x_{1}+x_{2}+x_{3}$ and $\tilde{\mu}=\xi$ (to have
the same notation as \cite{ber2}) we obtain the most general
solution of the Jacobi equation with $u=u_{0}+\xi, v=v_{0}+\xi,
w=w_{0}+\xi$ where $u_{0}=\mu\, \Psi_{3}, v_{0}=\mu \Psi_{2},
w_{0}=\mu \Psi_{1}$ as the known solution.

\vspace{0.3cm}

\noindent The above Lemma 4 suggests that all Poisson structures
in ${\mathbb R}^3$ have compatible pairs, because the condition
(\ref{con}) is not so restrictive on the Poisson matrices $J$ and
${\tilde J}$. Such compatible Poisson structures can be used to
construct bi-Hamiltonian systems.

\vspace{0.3cm}

\noindent {\bf Definition 2.} {\it A Hamiltonian equation is said
to be bi-Hamiltonian if it admits two Hamiltonian representations
$H$ and $\tilde{H}$, with compatible Poisson structures $J$ and
$\tilde{J}$ such that
\begin{equation}
\frac{dx}{dt}=J\nabla H=\tilde{J}\nabla\tilde{H}.
\end{equation}
} \vspace{0.3cm}

\noindent {\bf Lemma 4.} {\it Let $J$ be given by (\ref{GenSol})
and $H(x_1,x_2,x_3)$ is any differentiable function then the
Hamiltonian equation
\begin{equation}
\frac{dx}{dt}=J\nabla H =- \mu\, {\nabla \Psi} \times {\nabla H},
\end{equation}
is bi-Hamiltonian with the second structure given by $\tilde{J}$
with entries
\begin{equation}
\begin{array}{ccc}
\tilde{u}(x)&=&\tilde{\mu}\, \partial_1g(\Psi(x_1x_2x_3),H(x_1,x_2,x_3)),\\
\tilde{v}(x)&=&\tilde{\mu}\, \partial_2g(\Psi(x_1x_2x_3),H(x_1,x_2,x_3)),\\
\tilde{w}(x)&=&\tilde{\mu}\, \partial_3g(\Psi(x_1x_2x_3),H(x_1,x_2,x_3)),\\
\end{array}
\end{equation}
and $\tilde{H}=h(\Psi(x_1x_2x_3),H(x_1,x_2,x_3))$,
$\tilde{\Psi}=g(\Psi(x_{1},x_{2},x_{3}), H (x_{1},x_{2},x_{3}))$,
$\tilde{\mu}=\mu\frac{\partial_{\tilde{\Psi}}\Phi}{\partial_{\Psi}\Phi}$.
Provided that there exist differentiable functions $\Phi
(\Psi,{\tilde{\Psi}})$, $h(\Psi,H)$, and $g(\Psi,H)$ satisfying
the following equation
\begin{equation}\label{f2}
\frac{\partial g}{\partial \Psi} \frac{\partial h}{\partial H}
-\frac{\partial g}{\partial H}\frac{\partial h}{\partial
\Psi}=\frac{\Phi_1(\Psi,g)}{\Phi_2(\Psi,g)},
\end{equation}
where $\Phi_1=\partial_\Psi\Phi|_{(\Psi,g)}$,
$\Phi_2=\partial_{\tilde{\Psi}}\Phi|_{(\Psi,g)}$.}

\vspace{0.3cm}

\noindent{\bf Proof.} By Lemma~3, $J$ and $\tilde{J}$ are
compatible  and it can be shown by a straightforward calculation
that the equality (being a bi-Hamiltonian system)
\begin{equation}
\tilde{J}\nabla \tilde{H}=J\nabla H
\end{equation}
or
\begin{equation}
\tilde{\mu}\, \nabla\, \tilde{\Psi} \times \nabla \tilde{H}= \mu\,
\nabla \, \Psi \times \nabla\, H
\end{equation}

\noindent is guaranteed by (\ref{f2}). Hence the system
\begin{equation}
\begin{array}{lll}\label{sys}
\displaystyle{\frac{dx_1}{dt}}&=&\mu\partial_3\Psi\partial_2H
-\partial_2\Psi\partial_3H\\
\displaystyle{\frac{dx_2}{dt}}&=&-\mu\partial_3\Psi\partial_1H
+\partial_1\Psi\partial_3H\\
\displaystyle{\frac{dx_3}{dt}}&=&\mu\partial_2\Psi\partial_1H
-\partial_1\Psi\partial_2H\\
\end{array}
\end{equation}
is bi-Hamiltonian. $\Box$

\vspace{0.3cm}

\noindent {\bf Remark 5}.\, The Hamiltonian function $H$ is a
conserved quantity of the system. It is clear from the expression
(\ref{sys}) that the function $\Psi$ is another conserved quantity
of the system. Hence for a given Hamiltonian system there is a
duality between $H$ and $\Psi$. Such a duality arises naturally
because a simple solution of the equation (\ref{f2}) is
$\tilde{\Psi}=H$, $\tilde{H}=\Psi$ and $\tilde{\mu}=-\mu$.

\vspace{0.3cm}

\noindent Using the Lemma~4 we can construct infinitely many
compatible Hamiltonian representations by choosing functions
$\Phi,\, g,\, h$ satisfying (\ref{f2}). If we fix functions
$\Phi$, and $g$ then equation (\ref{f2}) became linear first order
partial differential equations for $h$. For instance, taking
$g=\Psi H$ and $\tilde{\mu}=-\mu$, which fixes $\Phi$, we obtain
$h=\ln H$. Thus we obtain second representation of equation
(\ref{sur}) with $\tilde{J}$ given by $\tilde{\Psi}=\Psi H$ and
$\tilde{H}=\ln H$.
 Let us give some examples of Hamiltonian systems. All the
autonomous examples in \cite{ber1} with bi-Hamiltonian
representations can be listed below but we prefer time dependent
ones.

\vspace{0.3cm}

\noindent {\bf Example 2.} Consider systems that are obtained from
RTW system \cite{PR}
\begin{equation}
\begin{array}{lll}
\dot{x}&=&\gamma x+\delta y+z-2y^2\\
\dot{y}&=&\gamma y-\delta x +2xy\\
\dot{z}&=&-2z(x+1).\\
\end{array}
\end{equation}
for appropriate subset of parameters by recalling. Following
\cite{GHHFC} we have:\\

\vspace{0.3cm}

\noindent {\bf RTW(1) system}
\begin{equation}
\begin{array}{lll}
\dot{x}_1&=&\delta x_2 +x_3e^{-2t}-2x_2^2\\
\dot{x}_2&=&-\delta x_1 +2x_1x_2\\
\dot{x}_3&=&-x_1x_3,\\
\end{array}
\end{equation}
where $\delta$ is an arbitrary constant. The matrix $J$ is given
by (\ref{GenSol}) with $\mu=1$ and
$\Psi=\displaystyle{\frac{1}{2}}(x_1^2-x_2^2+x_3e^{-t})$
and Hamiltonian $H=x_3(2x_2-\delta)$.\\
The matrix $\tilde{J}$ is given by (\ref{GenSol}) with
$\tilde{\mu}=1$ and
$\tilde{\Psi}=\displaystyle{\frac{x_3}{2}}(x_1^2-x_2^2+x_3e^{-t})(2x_2-\delta)$
and Hamiltonian
$\tilde{H}=\ln(x_3(2x_2-\delta))$.\\

\vspace{0.3cm}

\noindent {\bf RTW(3) system}
\begin{equation}
\begin{array}{lll}
\dot{x}_1&=&(x_3-2x_2)e^{-t}\\
\dot{x}_2&=&2x_1x_2e^{-t}\\
\dot{x}_3&=&-2x_1x_3e^{-t}.\\
\end{array}
\end{equation}
The matrix $J$ is given by with $\mu=1$ and
$\Psi=(x_1^2-x_2^2+x_3)e^{-t}$
and Hamiltonian $H=x_2x_3$.\\
The matrix $\tilde{J}$ is given by (\ref{GenSol}) with
$\tilde{\mu}=1$ and $\tilde{\Psi}=x_2x_3(x_1^2-x_2^2+x_3)e^{-t}$
and Hamiltonian
$\tilde{H}=\ln(x_3x_2)$.\\

\vspace{0.3cm}

\noindent {\bf RTW(4) system}
\begin{equation}
\begin{array}{lll}
\dot{x}_1&=&x_3e^{-(\gamma+2)t}-2x_2^2e^{\gamma t}\\
\dot{x}_2&=&2x_1x_2e^{\gamma t}\\
\dot{x}_3&=&-2x_1x_3e^{\gamma t},\\
\end{array}
\end{equation}
where $\gamma$ is an arbitrary constant. The matrix $J$ is given
by with $\mu=1$ and $\Psi=(x_1^2-x_2^2)+x_3)e^{\gamma t}
+x_3e^{-(\gamma+2)t}$
and Hamiltonian $H=x_2x_3$.\\
The matrix $\tilde{J}$ is given by (\ref{GenSol}) with
$\tilde{\mu}=1$ and
$\tilde{\Psi}=x_2x_3[(x_1^2-x_2^2)+x_3)e^{\gamma t}
+x_3e^{-(\gamma+2)t}]$ and Hamiltonian
$\tilde{H}=\ln(x_3x_2)$.\\

\vspace{0.3cm}

\noindent {\bf RTW(5) system}
\begin{equation}
\begin{array}{lll}
\dot{x}_1&=&\delta x_2+x_3-2x_2^2e^{-2t}\\
\dot{x}_2&=&-\delta x_1+2x_1x_2e^{-2t}\\
\dot{x}_3&=&-2x_1x_3e^{-2t},\\
\end{array}
\end{equation}
where $\delta$ is a non-vanishing constant. The matrix $J$ is
given by with $\mu=1$ and $\Psi=\displaystyle{\frac{\delta
e^{-2t}}{2}}(x_1^2-x_2^2)+ \displaystyle{\frac{\delta }{2}}x_3$
and Hamiltonian $H=x_1^2+x_2^2 +\displaystyle{\frac{2}{\delta }}x_2x_3$.\\
The matrix $\tilde{J}$ is given by (\ref{GenSol}) with
$\tilde{\mu}=1$ and $\tilde{\Psi}=[x_1^2+x_2^2
+\displaystyle{\frac{2}{\delta
}}x_2x_3][\displaystyle{\frac{\delta e^{-2t}}{2}}(x_1^2-x_2^2)+
\displaystyle{\frac{\delta }{2}}x_3]$ and Hamiltonian
$\tilde{H}=\ln(x_1^2+x_2^2 +\displaystyle{\frac{2}{\delta }}x_2x_3)$.\\

\vspace{0.3cm}

\noindent {\bf Remark 6}.\, \noindent If a system in ${\mathbb
R}^3$ is Hamiltonian then $\Psi$, and $H$ are constants of motion.
This reduces the system of three coupled ODE to a single ODE. As
an example let us consider Euler top equation given in Example 4.
The constants of motion are
$\Psi=-\frac{1}{2}(x_1^2+x_2^2+x_3^2)$, and
$H=\displaystyle{\frac{x^2_1}{2I_1}+\frac{x^2_2}{2I_2}+\frac{x^2_3}{2I_3}}$.
So, the equalities $\Psi=c_1$ and $H=c_2$ give (assuming $I_{1}
\ne I_{2} \ne I_{3}$)
\begin{equation}
\begin{array}{c}
x_1=\sqrt{\bar{c}_1 +\frac{I_1(I_3-I_2)}{I_3(I_2-I_1)}x_3^2} \\
x_2=\sqrt{\bar{c}_2 +\frac{I_2(I_3-I_1)}{I_3(I_1-I_2)}x_3^2} \\
\end{array}
\end{equation}
where $\bar{c}_{1}$ and $\bar{c}_{2}$ are new constants and $x_3$
is given by
\begin{equation}
t-t_{0}=\int \displaystyle {\frac{d x_3}{\sqrt{ \left (\bar{c}_1
+\frac{I_1(I_3-I_2)}{I_3(I_2-I_1)}x_3^2\right ) \left( \bar{c}_2
+\frac{I_2(I_3-I_1)}{I_3(I_1-I_2)}x_3^2 \right)}}}
\end{equation}

\noindent where $t_{0}$ is an integration constant. Hence one has
the general solution with the required number of integration
constants. \vspace{0.3cm}

This work is partially supported by the Turkish Academy of
Sciences and by the Scientific and Technical Research Council of
Turkey

\end{document}